\documentclass[a4paper]{jpconf}
\usepackage{graphicx}
\usepackage{hyperref}
\usepackage{amssymb}

\newcommand{\pT}{\mbox{$p_T$}}
\newcommand{\ET}{\mbox{$E_T$}}

\newcommand{\sqrts}{\mbox{$\sqrt{s}$}}

\newcommand{\jpsi}{\mbox{$J/\psi$}}
\newcommand{\psiprime}{\mbox{$\psi'$}}
\newcommand{\upsi}{\mbox{$\Upsilon$}}
\newcommand{\upsione}{\mbox{$\Upsilon(\textrm{1S})$}}
\newcommand{\upsitwo}{\mbox{$\Upsilon(\textrm{2S})$}}
\newcommand{\upsithree}{\mbox{$\Upsilon(\textrm{3S})$}}

\newcommand{\gevc}{\mbox{$\mathrm{GeV}/c$}}

\newcommand{\gevcc}{\mbox{$\mathrm{GeV}/c^2$}}

\newcommand{\pp}{\mbox{$pp$}}

\newcommand{\dAu}{\mbox{$d$+Au}}
\newcommand{\Npart}{\mbox{$N_{\mathrm{part}}$}}
\newcommand{\Ncoll}{\mbox{$N_{\mathrm{coll}}$}}

\newcommand{\bbbar}{\mbox{$b$-$\bar{b}$}}
\newcommand{\RAA}{\mbox{$R_{AA}$}}

\begin{document}
\title{Quarkonium Results in PbPb Collisions at CMS}

\author{M Calder\'on de la Barca S\'anchez$^1$, for the CMS Collaboration}

\address{$^1$ Physics Department, UC Davis. One Shields Avenue, Davis CA 95616, USA}

\ead{mcalderon@ucdavis.edu}

\begin{abstract}
We summarize the results from the study of charmonium and bottomonium via the dimuon
decay channel in PbPb collisions with the CMS experiment. 
We discuss the observation of sequential suppression of the \upsi\ states. 
We present preliminary results of prompt \jpsi\ and \psiprime\ production, 
as well as of non-prompt \jpsi's coming from the weak decay of $b$-quarks. 
This latter measurement is sensitive to $b$-quark energy loss. 
We discuss the results and compare to model predictions.
%
\end{abstract}

\section{Introduction}
Quarkonium suppression in a hot colored medium is a 
longstanding signature of quark-gluon plasma (QGP) formation.
It is expected that suppression will arise due to two main effects.  
The first mechanism proposed was color Debye screening \cite{matsui}. 
In the vacuum,
a heavy quark-antiquark bound state, such as charmonium or bottomonium, are kept together by the QCD color field.  Simulations of QCD
on the lattice show that the color fields take the form of a color-flux tube, which gives rise to an approximately linear potential.  Since the
quarks are massive and move at low velocities, one can use the framework of non-relativistic QCD. The spectroscopy of charmonia and bottomonia
can be obtained this way.  In a hot QCD medium, the expectation is that the additional color charges 
around the heavy quark-antiquark pair will
produce color fields which screen the potential between them.  The Debye screening radius depends 
on the temperature, and this then leads to 
a suppression of excited quarkonium states according to their average radius.  This picture has been 
augmented with recent studies which find that
there is also a broadening of the quarkonium spectral functions due to the interaction of 
soft gluons in the QGP medium with the quarkonium state.
The effects of gluo-dissociation and Landau damping can be encapsulated in the quark-antiquark 
potential as an imaginary part, with the real part 
governing the screening behavior.  For a review of this subject, see Ref.~\cite{Brambilla:2010cs}.  More recently, attempts 
to calculate the potential directly in the lattice have been
done, including the real and imaginary parts, i.e. screening and Landau damping/gluo-dissociation.
These studies motivate the continued study of
quarkonia in relativistic heavy ion collisions, since an observation of suppression could
then be connected to the properties studied in the lattice. In this
way, we could estimate the temperature reached in the medium. Furthermore, 
a suppression due to hot nuclear matter effects in the bottomonium case would
be a clean signature of deconfinement. (The picture is complicated in the case 
of charmonium due to the competing effect of recombination, which is
negligible for the case of bottomonium due to the small bottom 
cross section even at LHC energies.)
In this paper, we present a summary of some key results 
from the CMS in the field of quarkonium studies in relativistic heavy ion collisions.
The manuscript is organized as follows: We briefly describe the detector and analysis preliminaries,
then we discuss the results from prompt charmonium.  Next, we present the results
from non-prompt \jpsi\ which originate mainly from B meson decays. We then discuss
bottomonium measurements, followed by a summary of our findings.

\section{CMS Detector and Data}
A detailed description of the CMS detector can be found in Ref.~\cite{CMSdetectorChatrchyan:2008aa}. 
The main detectors used in this analysis are the Muon detectors 
and the silicon tracker. The solenoidal superconducting magnet 
of 6m internal diameter provides a $B=3.8$ T field. It houses
the silicon pixel and silicon strip tracker, as well as electromagnetic
and hadronic calorimeters. The tracker can measure charged tracks
in the kinematic range $|\eta|<$ 2.5. Muons are detected via
drift tubes, cathode strip chambers and resistive plate chambers 
in the range $|\eta|<$ 2.4. The granularity of the tracker working
in tandem with the strong bending $B$ field allow for measuring
muon transverse momenta (\pT) with resolution in the range $1-2\%$ for the
analyses presented here. The muons need a minimum total momentum in the range
$3 - 5$ \gevc\ to reach the muon stations.  This places limits on the acceptance
of \jpsi\ mesons. For mid-rapidity, we can measure \jpsi\ in the range
$\pT > 6.5$ \gevc, while at forward rapidity, we can measure it down to 
$\pT > 3$ \gevc.  For the \upsi\ mesons, the higher mass is transformed
to a higher muon momentum in the decay process, allowing us to measure 
\upsi\ mesons down to $\pT > 0$ for all rapidity.

To estimate the collision centrality, we measure the transverse
energy (\ET) in a hadron calorimeter placed in the forward region (HF,
$2.9 < |\eta| < 5.2$). We use a Glauber model to simulate the \Npart\ 
distributions based on fits to the HF \ET\ measurements, as described in Ref.~\cite{CMCcentralityChatrchyan:2011pb}.

\section{\jpsi\ Production}
In the invariant mass spectrum of dimuons, CMS can measure \jpsi\ mesons and estimate the prompt and non-prompt
contributions \cite{CMSJpsiPAS}.  We distinguish these two categories as follows. The prompt
component originates from directly produced \jpsi\ plus those which arise from the secondary decays 
of charmonium excited states, such as the \psiprime\ and $\chi_c$.  These excited state transitions
are sufficiently fast that the muons we observe are indistinguishable from primary muons.
The non-prompt component arises mainly from the decay of beauty hadrons, chiefly $B$ mesons.
These weak decays are longer lived, with lifetimes of order $c\tau\approx 490 \mu$m (for $B^+$).
By reconstructing the invariant mass, the position of the secondary vertex in the transverse plane $L_{xy}$, and the pair \pT, we
can do a simultaneous fit of the invariant mass spectrum together 
with the pseudo-decay length $\ell_{J/\psi}=L_{xy}m_{J/\psi}/\pT$.
This allows us to estimate the contributions of prompt and non-prompt \jpsi\ separately.

\subsection{Prompt \jpsi}
\begin{figure}[h]
\includegraphics[width=0.33\textwidth]{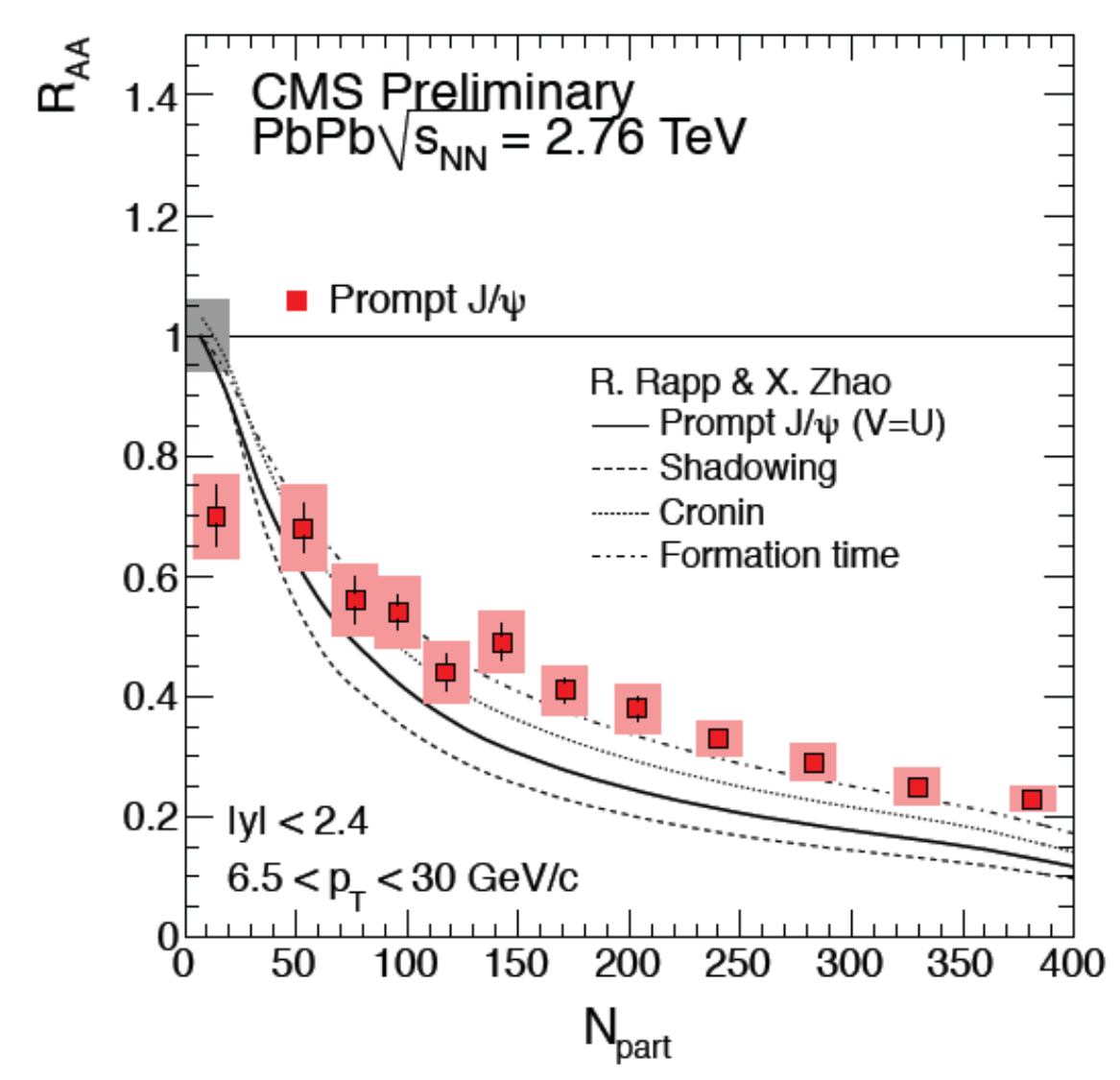}
\includegraphics[width=0.33\textwidth]{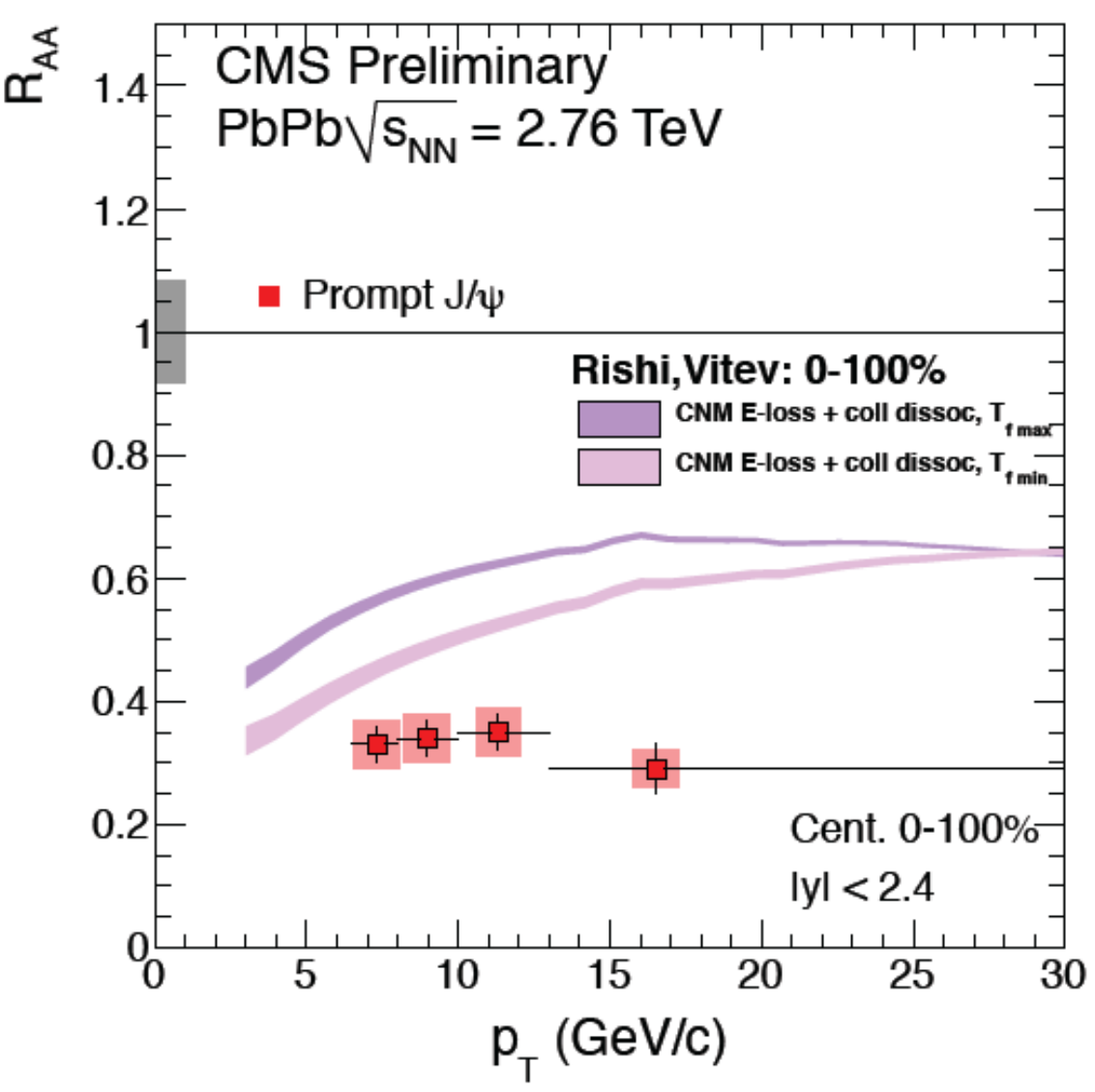}
\includegraphics[width=0.33\textwidth]{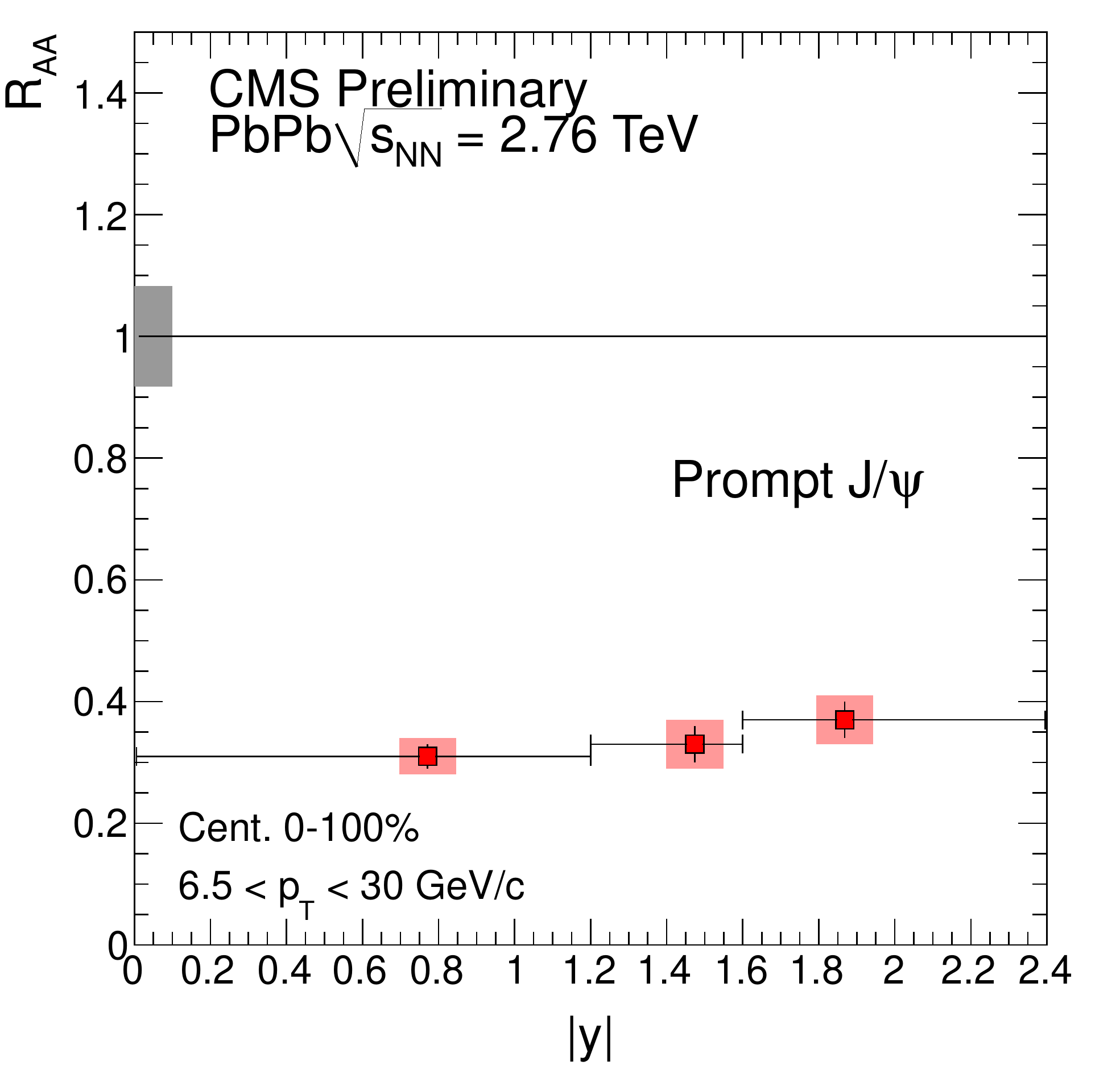}
\caption{\label{fig:PromptJpsiRAA}The nuclear modification factor,
\RAA, for prompt \jpsi\ as a function of \Npart\ (left), \pT (center) and $|y|$ (right). The
kinematic range for the measurement is $6.5 < \pT < 30$ \gevc, and $|y|<2.4$. 
The data show a clear suppression with increasing \Npart, but no \pT\ or $y$ dependence. For discussion
of the model comparisons, see text.}
\end{figure}

We have measured the production of prompt \jpsi\ in both PbPb and $pp$ collisions at \sqrts=2.76 TeV. 
To quantify the modifications observed in going from $pp$ to PbPb, we scale the yields measured in 
$pp$ using the Glauber model from Ref.~\cite{CMCcentralityChatrchyan:2011pb} and use this to normalize the PbPb measurements,
constructing the nuclear modification factor $R_{AA}$. Figure~\ref{fig:PromptJpsiRAA} shows $R_{AA}$ of prompt \jpsi
as a function of \Npart, \pT, and $y$.

For the top 5\% most central collisions in the kinematic range $\pT>6.5$ \gevc\ we observe a suppression of 
a factor of $\sim5$.  For the centrality-integrated data, we do not observe a significant \pT\ or rapidity
dependence of the suppression in our measured kinematics. Comparing the data to the model calculations
from Zhao and Rapp~\cite{Zhao:2011cv}, given by the various lines in the left panel of Fig.~\ref{fig:PromptJpsiRAA}, 
shows a qualitative agreement with the observed trend of
increasing suppression, i.e. decreasing \RAA, with increasing \Npart. In this model, for our kinematics,
the mechanism responsible for the suppression are the hot nuclear matter effects, screening and gluo-dissociation,
discussed previously. 

We see no evidence for a contribution from statistical recombination, which is expected to contribute
to the \jpsi\ yield at lower \pT. In contrast, a model from Sharma and Vitev~\cite{Sharma:2012dy} 
based on cold-nuclear-matter energy loss 
plus collisional dissociation of \jpsi\ mesons expects less suppression with increasing \pT,
in disagreement with the observed level of suppression that remains constant in our measurement
region. 

We have also measured the nuclear modification factor for the \psiprime\ meson.  We find
that in the kinematic region \pT\ $> 6.5 $ \gevc\ and $|y|<1.6$ the \psiprime\
 is suppressed much more than the \jpsi. Our preliminary result in this kinematic 
 region integrated over all centralities is 
 $\RAA (\psi(2S)) = 0.11 \pm 0.03\ (\mathrm{stat.}) \pm 0.02 (\mathrm{syst.}) \pm 0.02 (pp)$.  
 In the forward
rapidity region, $1.6 < |y| < 2.4$ and lower \pT\ (down to 3 \gevc) we find an enhancement of
the \psiprime. However, the current statistics from the \pp\ measurement 
hinder the statistical significance of the forward $y$ measurement. With \pp\ data from 2013 this
situation will be ameliorated.

\subsection{Non-prompt \jpsi\ production}
\begin{figure}[h]
\includegraphics[width=0.33\textwidth]{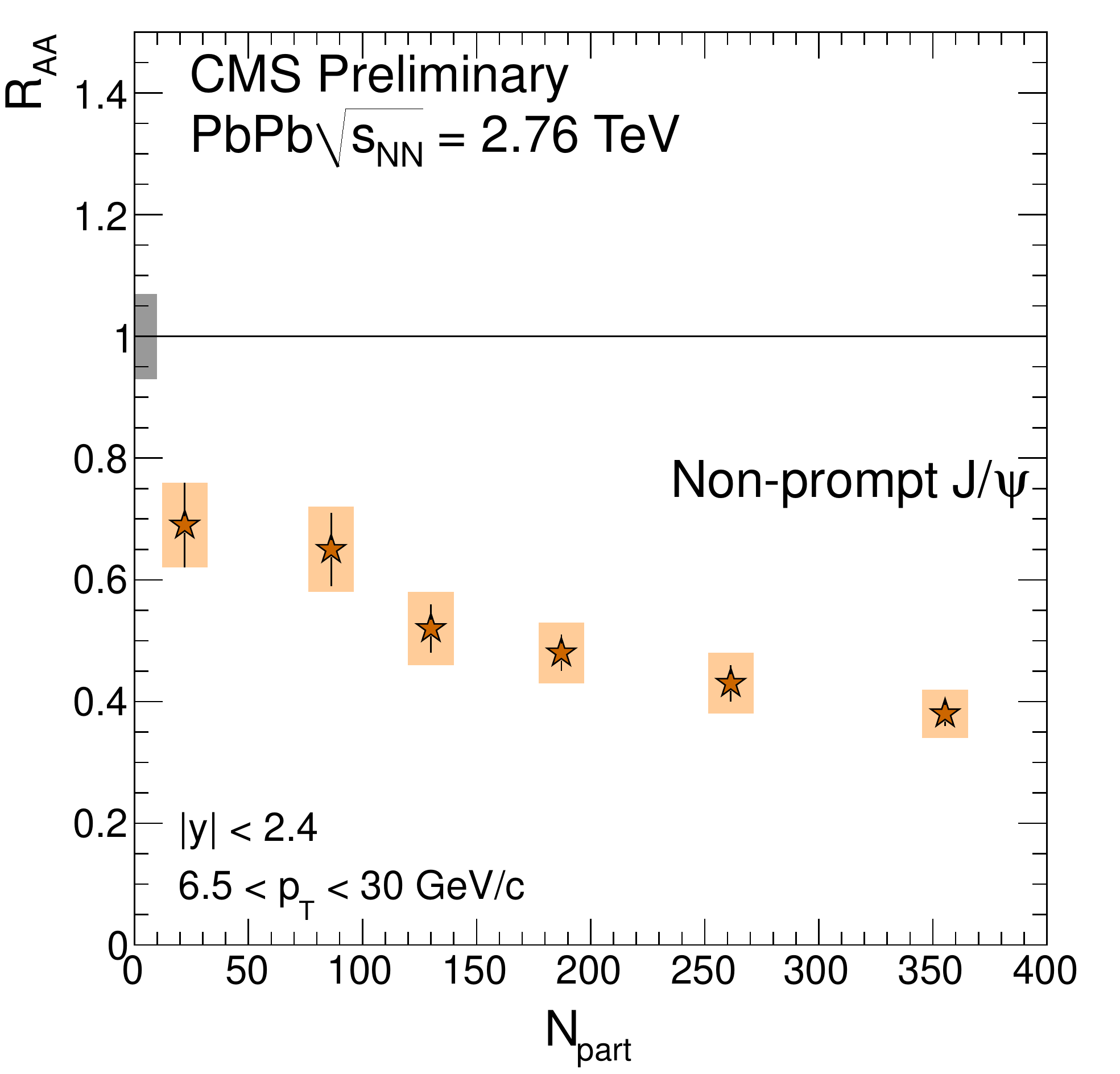}
\includegraphics[width=0.33\textwidth]{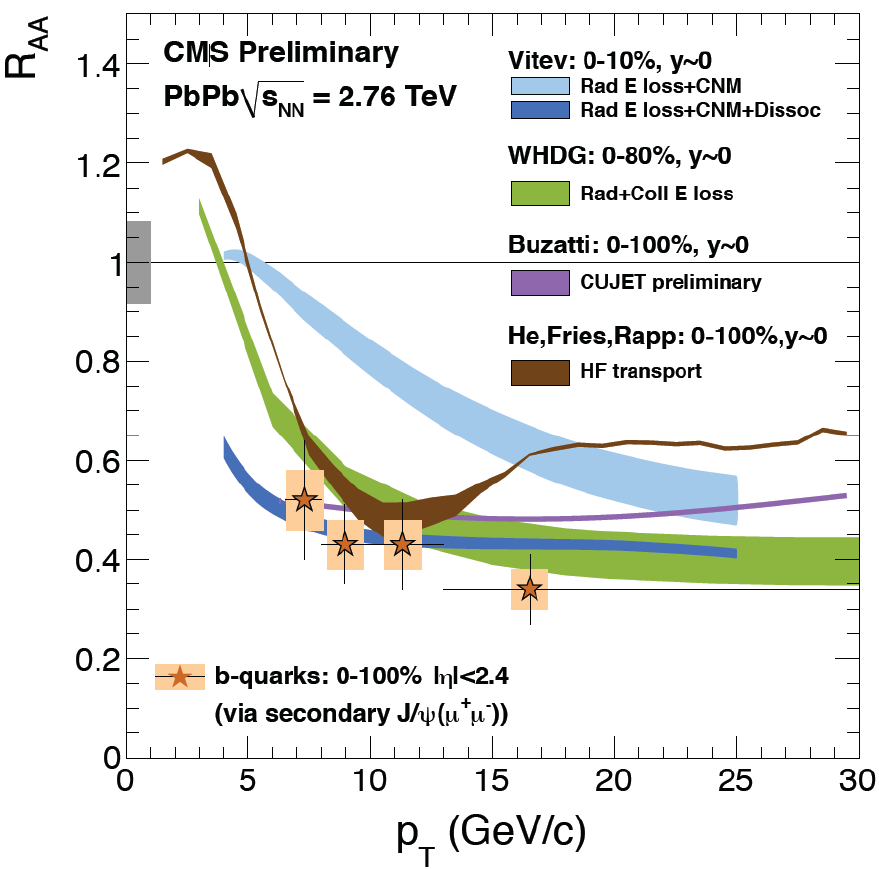}
\includegraphics[width=0.33\textwidth]{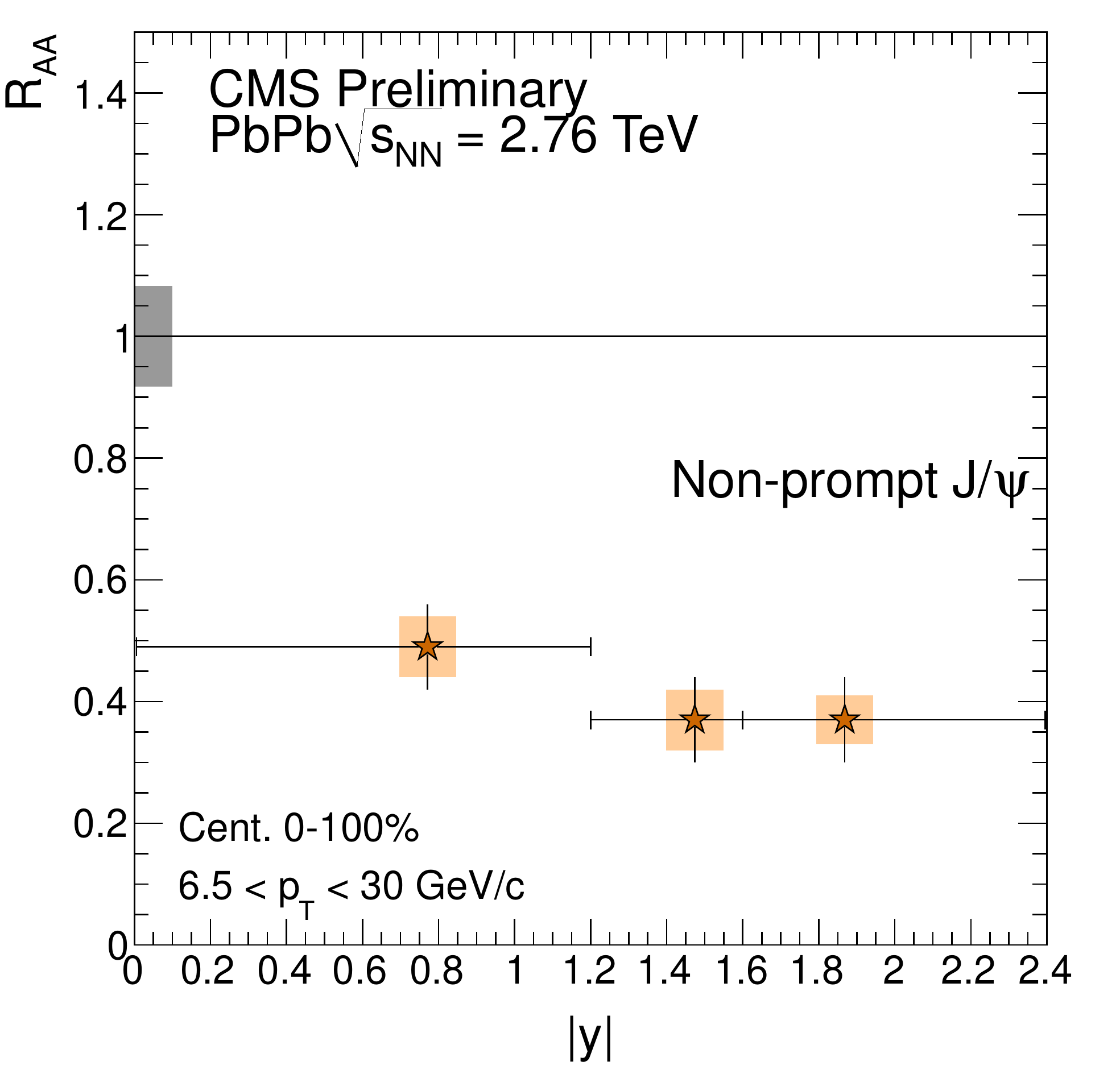}
\caption{\label{fig:NonPromptJpsiRAA}The nuclear modification factor,
\RAA, for non-prompt \jpsi, which mainly come from B meson decays, 
as a function of \Npart\ (left), \pT (center) and $|y|$ (right). The
kinematic range for the measurement is $6.5 < \pT < 30$ \gevc, and $|y|<2.4$. 
For discussion of the experiment and model comparisons, see text.}
\end{figure}

The non-prompt contribution to the inclusive \jpsi\ yield was also extracted from the PbPb and $pp$ data.
The nuclear modification factor \RAA\ is shown in Figure~\ref{fig:NonPromptJpsiRAA}.
The dependence of \RAA\ on centrality, \pT\ and $y$ is shown in the left, middle and right panel, respectively.

We find an increasing suppression for non-prompt \jpsi\ with increasing \Npart. 
The suppression reaches a factor 2.5 ($\RAA \approx 0.4$) for the top 5\% most central events.
When integrated over centrality, there is a hint of increasing suppression with increasing \pT\ or $y$.
As mentioned earlier, the non-prompt \jpsi\ production is dominated by the decays of $B$ mesons.
Therefore, suppression of the non-prompt \jpsi\ yield is expected to be a sensitive probe to 
energy loss of $b$-quarks.  Due to the large mass of the $b$ ($m_{b}\sim 4.2$ \gevcc), the radiative
energy loss of $b$ quarks is expected to be suppressed, when compared to light parton gluon 
Bremsstrahlung, at forward angles of order $\theta\lesssim m/E$, where $\theta$ is the angle
between the radiated gluon and the direction of the propagating quark, $m$ is the quark mass
of the heavy quark, and $E$ its Energy. Measurements of non-photonic electron \RAA\ at RHIC
indicated that, in addition to radiative energy loss, $b$ quark production also had a
significant modification due to collisional energy loss.   In the comparison of the \RAA\ vs.~\pT\ 
data to model calculations, we find that the model from Vitev, based on Ref.~\cite{Vitev:2008jh}, with only 
radiative energy loss does not give enough suppression to be consistent with our measurement (light blue
band). The addition of collisional energy loss (dark blue) is in much better quantitative agreement
with our data.  A calculation in the energy-loss framework developed by 
Wicks, Horowitz, Djordjevic and Gyulassy, including both the radiative and collisional
energy-loss mechanisms, is also shown. The framework was developed for comparing to RHIC data,
and is based on pQCD weak-coupling and AdS/CFT strong-coupling drag energy loss models \cite{HorowitzPANIC2011}. 
The model parameters were set at RHIC and then kept constant to predict the \RAA\ at LHC (green band).
We see that this model is in good quantitative agreement with our data. Note that our data are given in terms of
the \pT\ of the measured \jpsi\, while the calculations are given in terms of the \pT\ of the
$B$ meson, which is always higher than the daughter \jpsi\ \pT. Therefore, one would expect a slight
shift of the model curves to lower \pT\ to account for this. We also compared our data to calculations
from the CUJET model \cite{Buzzatti:2012pe} and He et al. \cite{He:2011qa}, which also are in
good agreement with our data at low \pT. The energy-loss model from He et al., which is based on a heavy-quark
diffusion and hadronization picture, expects less suppression with increasing \pT\, which is in disagreement with our
highest measured \pT\ \RAA.

\section{Bottomonium}
\begin{figure}[h]
\includegraphics[width=0.33\textwidth]{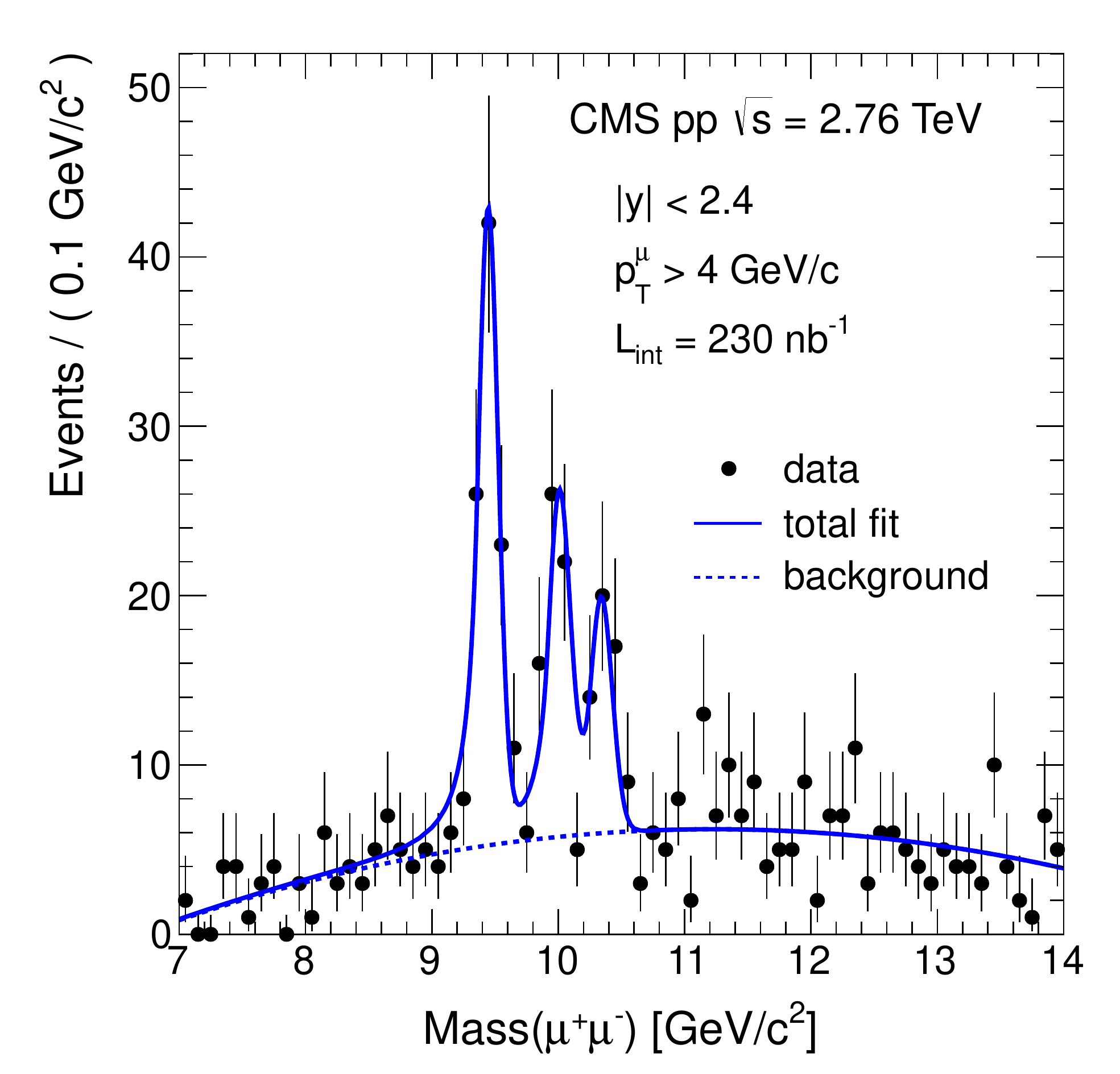}
\includegraphics[width=0.33\textwidth]{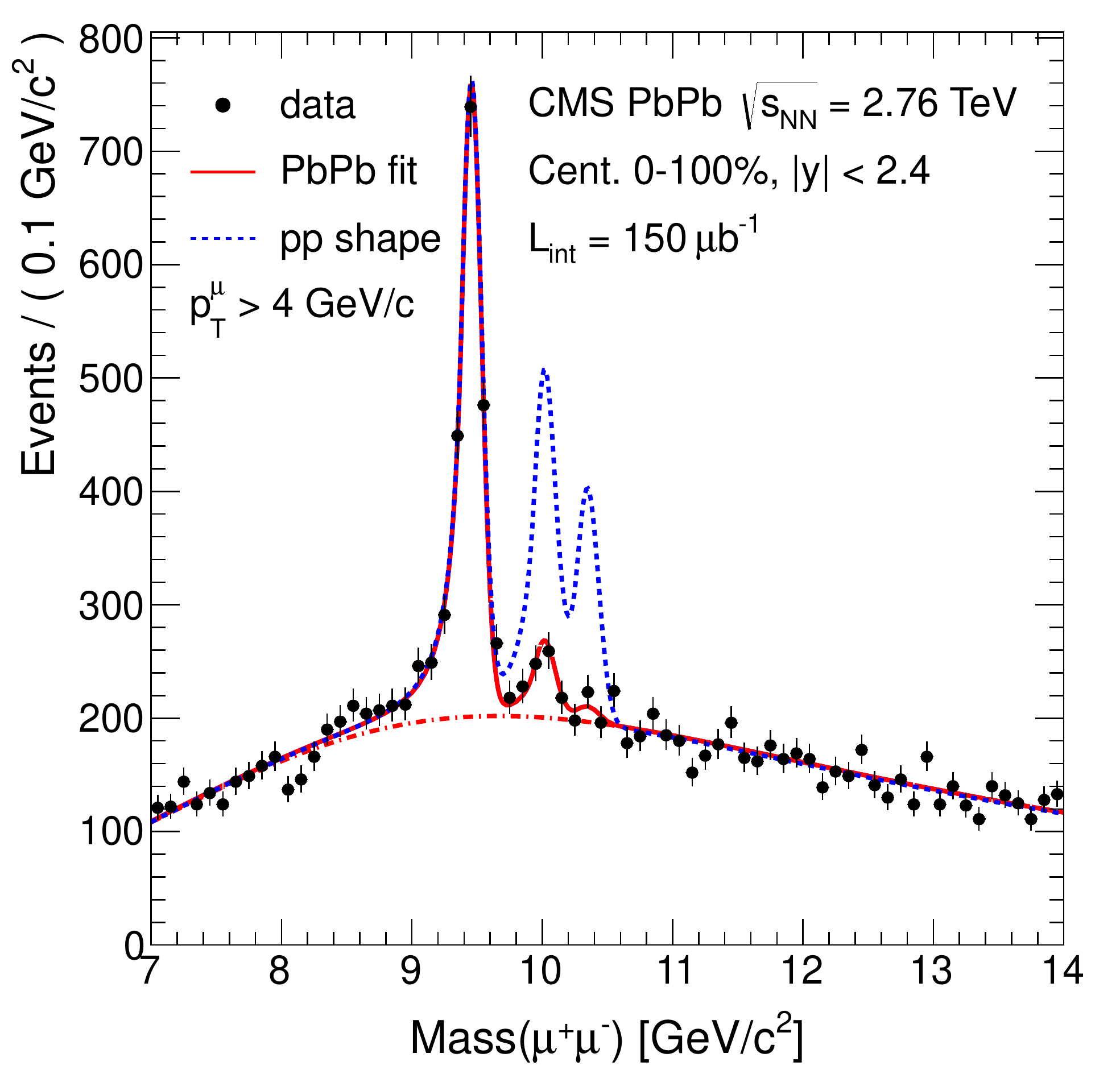}
\includegraphics[width=0.33\textwidth]{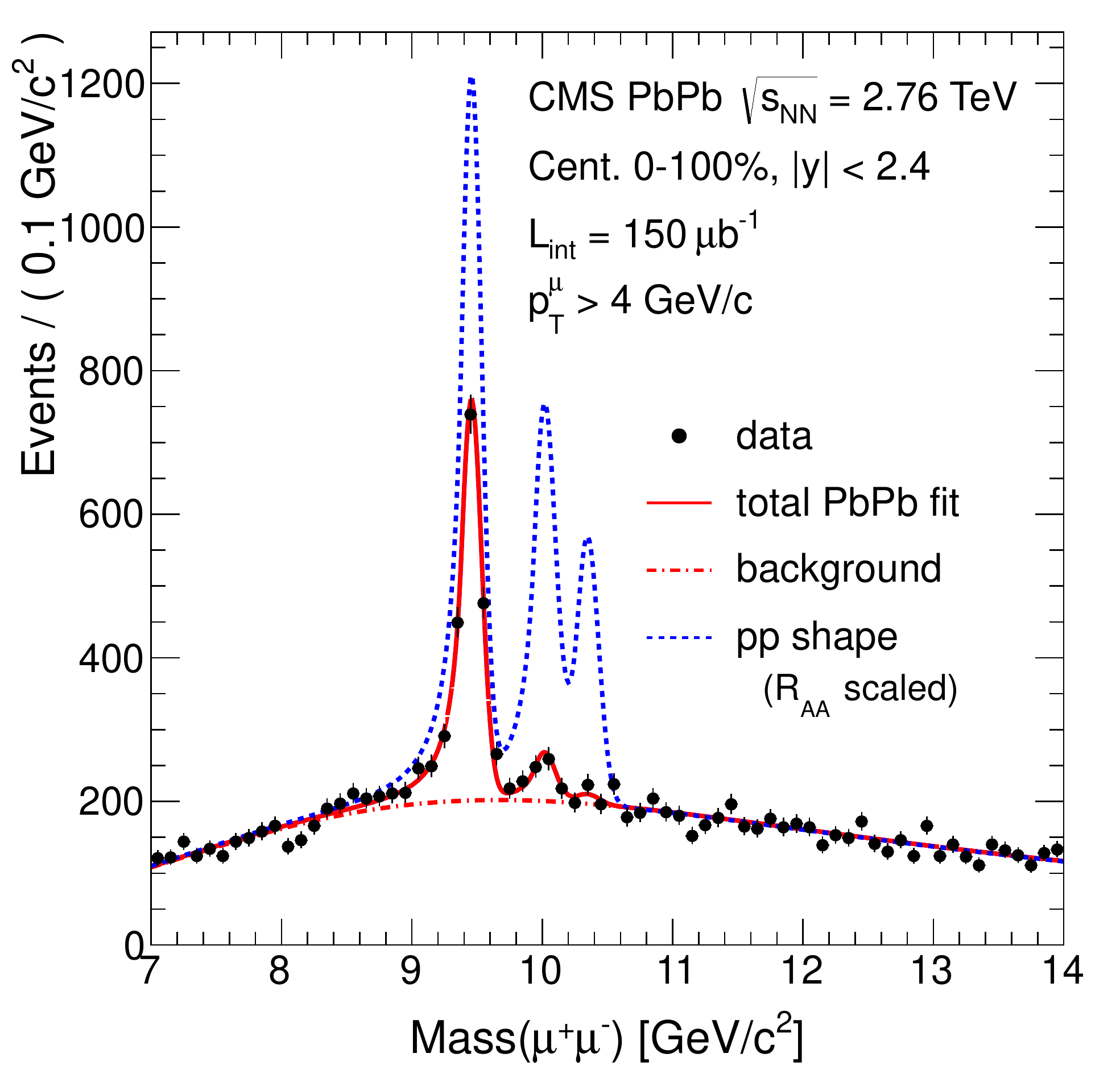}
\caption{\label{fig:UpsilonInvMass} The invariant mass distribution of dimuons.
The $pp$ data are shown in the left panel. The center and right panel both show the PbPb data.
The $pp$ and PbPb data are fit, with the result shown on the plot.  The $pp$ fit is shown in the
center and right panels to graphically illustrate the double ratio 
$(\upsi(nS)/\upsi(1S))_{\mathrm{PbPb}}/(\upsi(nS)/\upsi(1S))_{pp}$ (center) and
the nuclear modification factor scaling \RAA\ (right). See text for details.}
\end{figure}

We have measured the \upsi\ family in both the $pp$ and PbPb collision systems.  Figure~\ref{fig:UpsilonInvMass}
shows the invariant mass of dimuon pairs measured in $pp$ (left) and PbPb (center and right).
The excellent resolution of the CMS detector for muons allows a clear separation
of all 3 states \cite{Chatrchyan:2012lxa}. The data are fit in order to extract various parameters of interest, in particular
the raw (not acceptance corrected) single ratios $\upsi(nS)/\upsi(1S)$ in both collision systems and
the double ratio of these raw single ratios between the PbPb and $pp$ system 
$(\upsi(nS)/\upsi(1S))_{\mathrm{PbPb}}/(\upsi(nS)/\upsi(1S))_{pp}$.
The raw single ratios cancel uncertainties arising from mechanisms which affect the excited
states and the ground state in the same way. An example of such a mechanism is the
uncertainty arising from the parton distribution functions. For \upsi\ production,
the dominant production channel is gluon-gluon fusion, hence the gluon PDF is the
one that plays the most important role.  Since this PDF enters in the same way in the
production of the bottomonium final state, namely in the initial state entrance channel,
any modifications will affect the ground state and the excited states in the same way, thus 
they will cancel in the single ratio.  An important experimental effect which does 
not cancel in the raw single ratio is a difference in the acceptance for excited states compared to 
the ground state.  This difference in acceptance arises from the fact that the
muons chosen for the analysis must satisfy $\pT^\mu > 4$ \gevc.  Since we look for two muons in
order to reconstruct the \upsi\ mesons, and the \pT\ threshold is close to half of the \upsi\ mass,
it is appreciably easier for a state at higher mass to have its muon daughters
satisfy the \pT\ threshold compared to a lower-mass \upsi\ meson. This acceptance effect will
cancel in the double ratio between PbPb and $pp$.  This makes the double ratio a very robust
observable for quantifying modifications of the excited states with a very small systematic 
uncertainty.  This is illustrated graphically in the center panel of Fig.~\ref{fig:UpsilonInvMass}.
The shape of the invariant mass fit to the \upsi\ states in the $pp$ collision system is overlaid
on the PbPb data. In the single ratio, the \upsi(1S) yield is the reference, and in the double ratio 
the PbPb and pp ratios are then compared. By construction, the double ratio for the 1S
is 1, so the 1S peaks in PbPb and $pp$ have the same height in the figure. The interesting
observation then is to see what happens to the excited states. If the excited-to-ground state
ratio in PbPb was the same as in $pp$, the blue dotted line ($pp$ shape) would also match the 
PbPb data for the excited states. Clearly it does not.  Therefore, we conclude that the excited states
are modified in PbPb. Furthermore, we surmise that this modification 
is not arising from nuclear PDFs, because they would affect the excited states in the same
way as the ground state.
It remains to be seen whether this is strictly a hot nuclear matter effect by doing this measurement
in the pPb collision system.

In the right panel, we illustrate graphically a different scaling.  To compare absolute yields for each
state between $pp$ and PbPb, we again use a Glauber model to estimate the nuclear overlap and the
number of collisions, \Ncoll.  Using this information, we scale the measured yields in $pp$ and compare
them directly to the data. The blue dotted line in the right panel shows this expected $pp$ yield, which is
the denominator in the construction of the nuclear modification factor \RAA.  One can see in this 
way that the inclusive yield for all \upsi\ states in PbPb collisions shows suppression compared to 
the binary-scaled $pp$ yield.

\begin{figure}[h]
\includegraphics[width=0.33\textwidth]{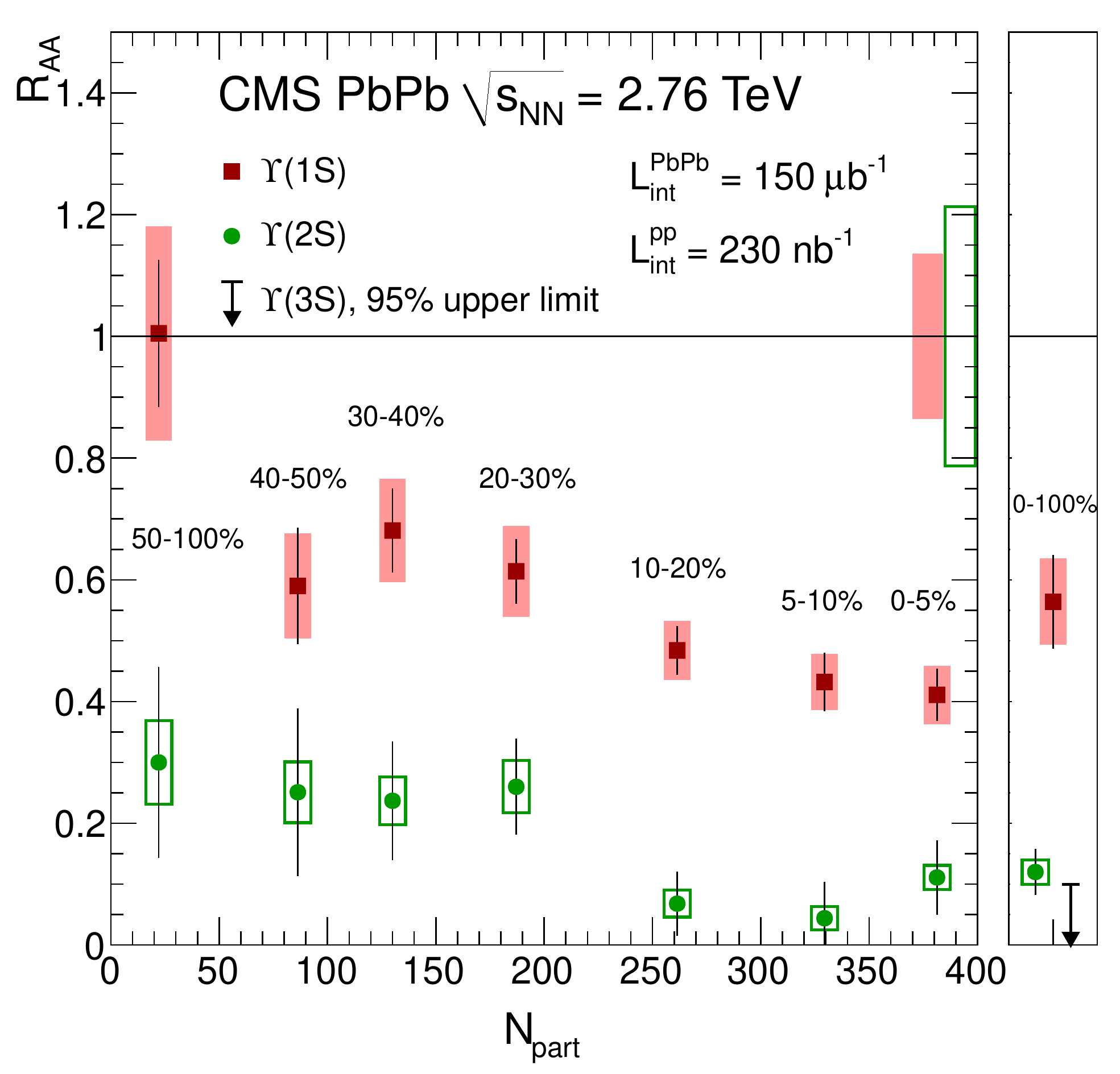}
\includegraphics[width=0.33\textwidth]{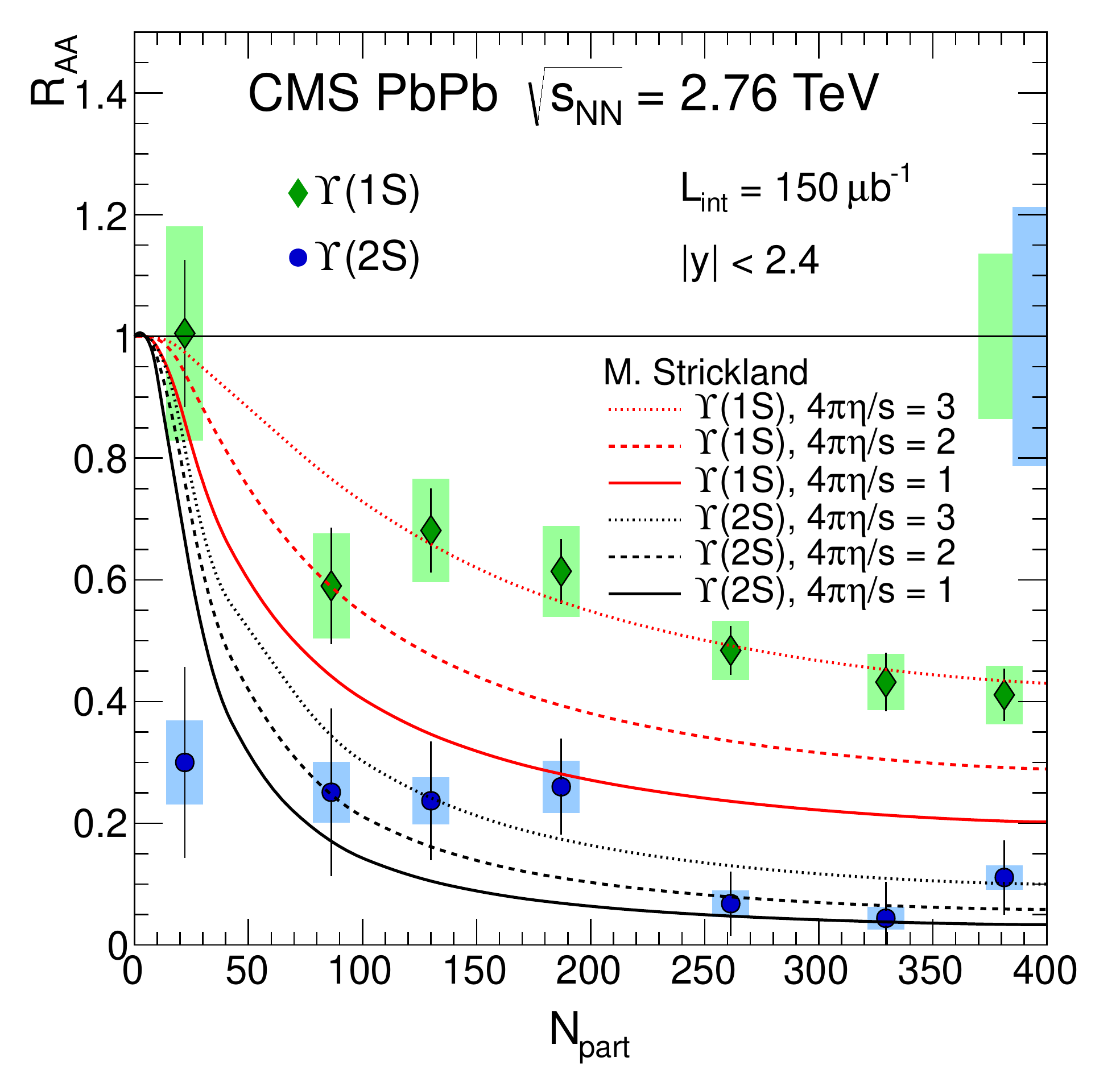}
\includegraphics[width=0.33\textwidth]{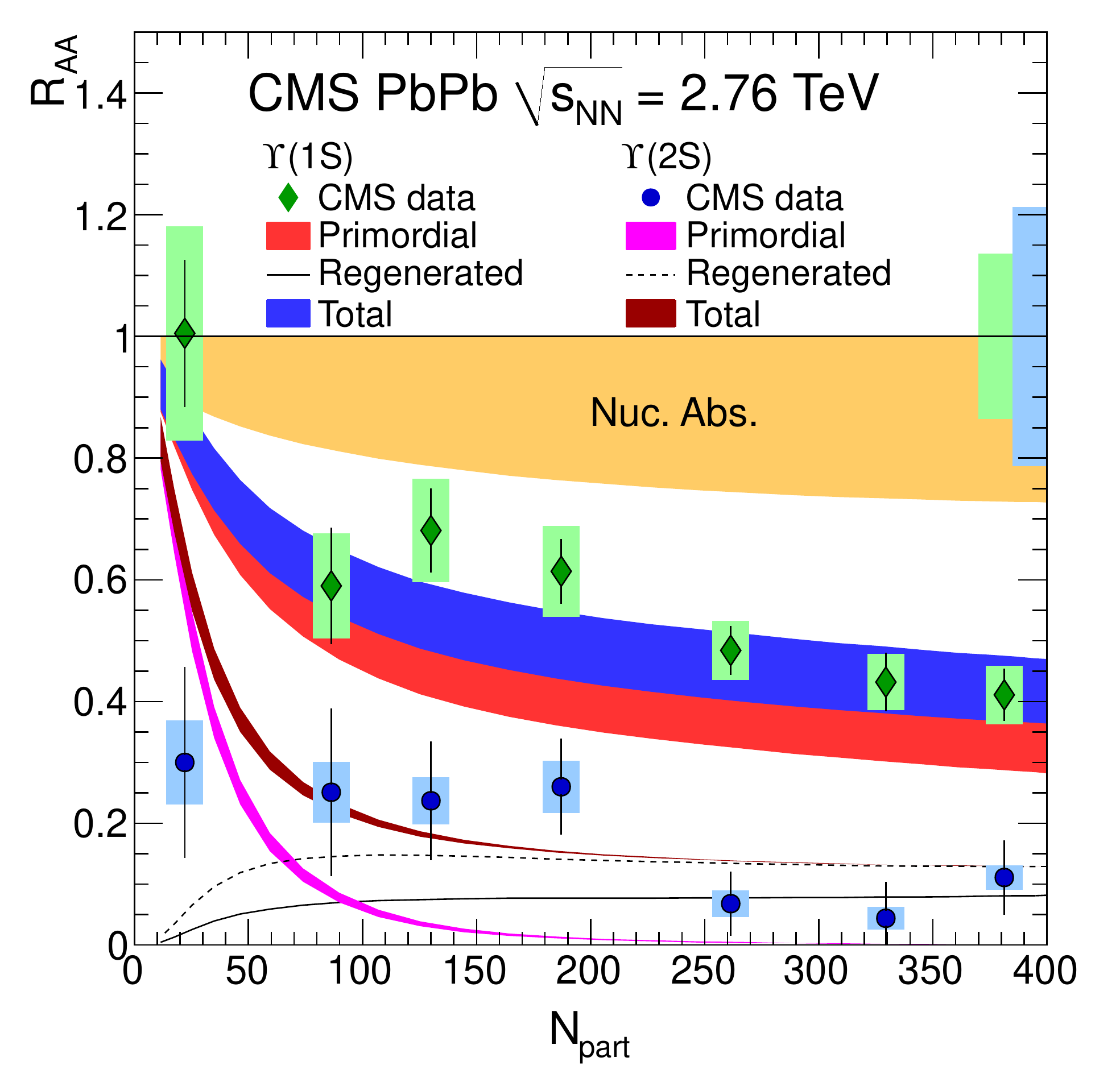}
\caption{\label{fig:UpsilonRAA} The nuclear modification factor, \RAA, for \upsi\ mesons 
as a function of \Npart. (Left) The red filled squares show our results for \upsione\ 
and the green filled circles are for \upsitwo.  The small panel to the right of this figure shows
the centrality integrated values.  The upper limit for the \upsithree\ is shown as the arrow 
in this panel.  (Center) The same data for the \upsione\ and \upsitwo\ are shown in
as green diamonds and blue circles, respectively. The data are compared to a model calculation
from Ref.~\cite{Strickland:2011aa}.  (Right) The \upsi\ data are the same as in the center panel. The data
are compared to a model calculation from Ref.~\cite{Emerick:2011xu}. See text for details.}
\end{figure}

The results for the nuclear modification factor, \RAA\ of \upsi\ mesons are shown in Fig.~\ref{fig:UpsilonRAA}.
With the 150 $\mu b^{-1}$ of integrated luminosity sampled in 2011, we are able to split the 
\upsione\ and \upsitwo\ data into 7 centrality bins.  The \upsione\ data are shown as the red
filled squares in the left panel of Fig.~\ref{fig:UpsilonRAA}, and as the green diamonds
in the center and right panels.  The \upsitwo\ data are shown as the green filled circles
in the left panel, and as the blue filled circles in the center and right panels.  The small panel to the
right of the left panel shows the centrality integrated \RAA\ values:
\begin{itemize}
\item \RAA(\upsione) $ = 0.56 \pm 0.08 (\mathrm{stat.}) \pm 0.07 (\mathrm{syst.})$
\item \RAA(\upsitwo) $ = 0.12 \pm 0.04 (\mathrm{stat.}) \pm 0.02 (\mathrm{syst.})$
\item \RAA(\upsithree) $ < 0.1 $ at 95\% C.L.
\end{itemize}
This is the first measurement of \upsitwo\ \RAA.  In addition, the \RAA\ values we observe are 
ordered in the expected way: smaller \RAA\ for the states with the smaller binding energy and larger
radius.

In studying the data as a function of centrality, we find that the \upsione\ and \upsitwo\ suppression
increases with \Npart.  For all the centrality bins, 
we observe the suppression of the \upsitwo\ state to 
be larger than the suppression of the \upsione.  In the most peripheral bin, the \upsione\ nuclear
modification factor is consistent with unity, while that for the \upsitwo\ remains low.  However,
it should be noted that this bin includes a wide impact parameter range (50-100\% of the total
cross section), and it is expected that most of the events where an \upsi\ will be produced will be 
biased towards larger \Ncoll\ and hence smaller impact parameter. 

Our measurements are of inclusive production of \upsi\ mesons, which include both the directly
produced \upsi\ mesons and those which are the daughters of excited states. For example,
the $\chi_b$ meson can decay via $\chi_b \rightarrow \upsione + \gamma$.   
The $\chi_b$ is less tightly bound and has a larger average radius than the \upsione.  If the
$\chi_b$ states are suppressed in the QGP, this will reduce the contribution from
one of the sources of \upsione\ and will lead to a suppression even if the directly produced
\upsione\ mesons do not melt in the QGP or are otherwise modified.  This contribution
from feed-down of higher excited states has been measured by
CDF at lower \sqrts\ and by LHCb at higher \sqrts\ \cite{Affolder:1999wm,Aaij:2012se}.  Both experiments find
that the feed-down contribution to the \upsione\ yield is $\approx50\%$.  However,
this contribution has only been measured at high \pT.  If the contribution is the same at lower \pT, a suppression
of all the excited states down to \RAA\ = 0 but with no suppression of \upsione\ would result in 
$\RAA \approx 0.5$ for the \upsione.  Under this hypothesis, our data for the most central
events for the \upsione\ is consistent with the suppression affecting
only the excited states.  This scenario illustrates the importance of including the effects of feed-down
when making comparisons to our inclusive measurements. 

Two such comparisons are shown in the center and right panels of Fig.~\ref{fig:UpsilonRAA}. In the center
panel, we compare our results to a model calculation by Strickland and Bazow \cite{Strickland:2011aa}.
In this model, the heavy-quark potential is taken to be the heavy-quark internal energy as calculated in
the lattice.  (The calculation was also done with the free energy and found to be inconsistent with our data. The
free-energy calculation is not shown.)  In addition, the model also includes the imaginary part of the 
potential (the Landau damping and gluo-dissociation effects discussed previously). These ingredients 
naturally lead to sequential melting.  In addition to including the effects of feed-down, 
under the assumption that the contribution from excited states is as measured by CDF 
in the entire \pT\ range, the model includes a dynamical expansion using
the framework of anisotropic hydrodynamics.  The model was run under three different choices for the value
of the shear viscosity to entropy ratio, $4\pi \eta/S = $1, 2, and 3.  For each of these choices, the initial
temperature was also adjusted in order to match the existing measurements of multiplicity. The
temperatures used were in the range 552 $< T_0 < $ 580 MeV. The highest temperature corresponds
to the lowest shear viscosity to entropy ratio, and these are shown as the solid lines in Fig.~\ref{fig:UpsilonRAA},
center.  The red line is for the \upsione\ and the blue line is for the \upsitwo.
The \upsione\ data seems to be best described by the calculation with the lowest temperature
and highest shear viscosity to entropy ratio in this model. There is some tension to describe 
the \upsitwo\ data with this same set of parameters, although with the statistical uncertainty 
for the \upsitwo\ data none of the choices can be ruled out to more than 2$\sigma$.

The right panel of Fig.~\ref{fig:UpsilonRAA} shows a comparison to a model calculation
from Emerick et al.~\cite{Emerick:2011xu}.  The authors of this calculation
considered two scenarios. In one scenario, the binding energy of the states
was varied with temperature, such that at large temperature the binding
was small.  This was dubbed the ``Weak-Binding Scenario''. The second scenario
assumed that there is no variation of the binding energy with temperature. 
The bound states remain with the same masses and binding energies
as in the vacuum throughout their dynamical evolution in the QGP. This
was called the ``Strong-Binding Scenario''. As an example of the
differences, the bottomonium spectral function 
is much wider in the Weak-Binding compared to the Strong-Binding scenario.
In comparing this to lattice data using the so-called euclidean correlator,
which is the integral of the spectral function over some kernel, 
it was found that the Strong-Binding scenario results in a better
agreement with the finding that the correlators do not change very much
with temperature.  This has been taken as a strong motivation for 
preferring the Strong-Binding scenario. In this model, both Landau
damping and gluo-dissociation become the leading break-up mechanisms,
with the former being more important for weakly bound states and the
latter for states with large binding energy.  The 
dynamical evolution is done in the framework of kinetic theory.
The primordial suppression, including feed-down contributions, is shown
as the red band in Fig.~\ref{fig:UpsilonRAA}, right, for the \upsione.  For
the \upsitwo\ it is shown as the magenta band. In this model, the primordial 
\upsitwo\ yield is completely suppressed in the most central collisions.  
This model also includes a
parameterization of cold-nuclear matter effects, based on preliminary
results from \dAu\ collisions from STAR \cite{Reed:2011zza}.  
In addition, the model incorporates
a contribution from regeneration of \upsi\ mesons via the coalescence of
uncorrelated \bbbar\ pairs produced in the same event. The solid black line
shows this contribution for the \upsione, and the dotted black line
shows the corresponding contribution for the \upsitwo.  The total
nuclear modification factor, including all contributions, is shown 
as the blue band for the \upsione\ and as a brown band for the
\upsitwo.  Given the small \bbbar\ cross section, the total 
suppression is not too different from the primordial result in this model.
There is still a large uncertainty in the nuclear absorption, underlining
the need for results from \dAu\ and $p$+Pb collisions.
In this model the initial temperature reached in the QGP is
$T_0 \approx 610 $ MeV.

\section{Summary}
\begin{figure}[h]
\centering
\includegraphics[width=0.45\textwidth]{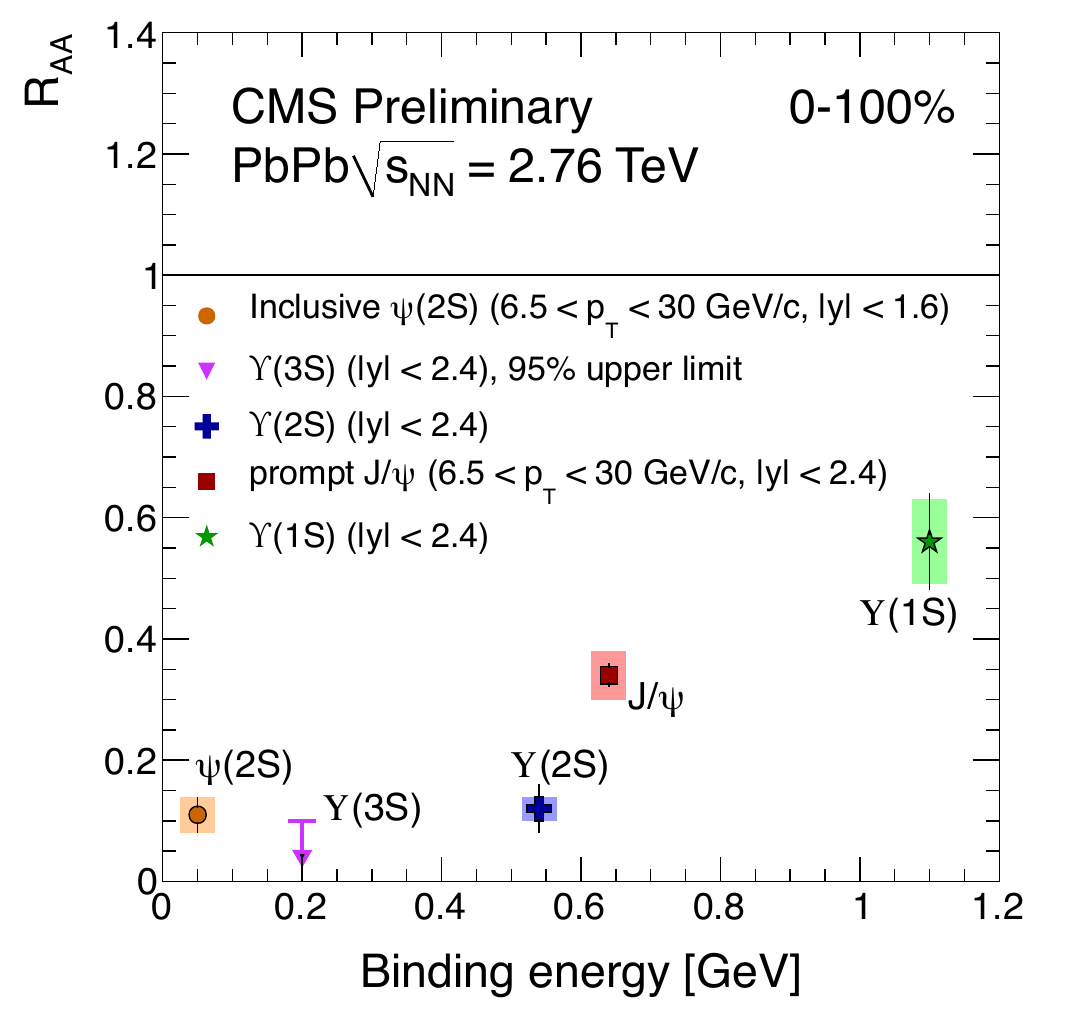}\hspace{2pc}
\begin{minipage}[b]{14pc}\caption{\label{fig:RAABindingEnergy} 
Nuclear modification factor, \RAA, for 
the quarkonia measured by CMS in heavy ion collisions shown as a function
of the binding energy of the state.   A sequential suppression pattern is observed.
See text for details.}
\end{minipage}
\end{figure}

We presented the results from Quarkonia production measured by CMS in the
heavy ion collision environment. We have measured charmonia at high-\pT\ 
in the mid-rapidity region.  We found that the \jpsi\ mesons are suppressed, 
and the \psiprime\ mesons at mid-rapidity are suppressed more than the \jpsi.
This pattern is as expected from the sequential melting picture.
The non-prompt \jpsi\ measurement allows us to study $b$-quark energy loss.
Calculations incorporating radiative and collisional energy loss are
consistent with our preliminary measurements.
In the bottomonium sector, we also observe a clear ordering of the suppression
of the three \upsi\ states with binding energy.  A plot summarizing
the observed nuclear modification factors for all the quarkonia states
discussed in this manuscript is shown in Fig.~\ref{fig:RAABindingEnergy}.
The horizontal axis indicates the binding energy of the given quarkonium
state. The \psiprime\ is the state with the smallest binding energy, and
the \upsione\ is the most tightly-bound state.  The binding energy is simply taken to be
the vacuum binding energy, i.e. the mass difference between the quarkonium
state and the open heavy-flavor continuum, given by twice the mass of the $D$ meson
for the charmonium states and twice the mass of the $B$ meson for the bottomonium
family.  We observe a pattern of sequential suppression of quarkonium
states.  This pattern is consistent with the expectation that quarkonia will
melt in a hot quark-gluon plasma, with the least tightly bound states melting first.
The temperatures reached in the plasma in the two model calculations used for comparison
to our data are in the range 550 -- 610 MeV.  We look forward to measurements of quarkonia
in $p$+Pb collisions, which will be important to distinguish cold and hot nuclear matter effects.


\section*{References}


\begin{thebibliography}{9}
\bibitem{matsui} T.~Matsui and H.~Satz, Phys. Lett. B {\bf 178}, 416 (1986)

\bibitem{Brambilla:2010cs} 
  N.~Brambilla, S.~Eidelman, B.~K.~Heltsley, R.~Vogt, G.~T.~Bodwin, E.~Eichten, A.~D.~Frawley and A.~B.~Meyer {\it et al.},
  Eur.\ Phys.\ J.\ C {\bf 71}, 1534 (2011)
  [arXiv:1010.5827 [hep-ph]].

\bibitem{CMSdetectorChatrchyan:2008aa}
  S.~Chatrchyan {\it et al.}  [CMS Collaboration],
  JINST {\bf 3}, S08004 (2008).
  
\bibitem{CMCcentralityChatrchyan:2011pb} 
  S.~Chatrchyan {\it et al.}  [CMS Collaboration],
  JHEP {\bf 1108}, 141 (2011)
  [arXiv:1107.4800 [nucl-ex]].
  
 \bibitem{CMSJpsiPAS}
 CMS  Collaboration, \jpsi\ results from CMS in PbPb collisions with 150$\mu$b$^{-1}$ data, \href{http://cds.cern.ch/record/1472735?ln=en}{PAS HIN-12-014.}
 
\bibitem{Zhao:2011cv} 
  X.~Zhao and R.~Rapp,
  Nucl.\ Phys.\ A {\bf 859}, 114 (2011)
  [arXiv:1102.2194 [hep-ph]].

\bibitem{Sharma:2012dy} 
  R.~Sharma and I.~Vitev,
  Phys.\  Rev.\  C 87, {\bf 044905} (2013)
  [arXiv:1203.0329 [hep-ph]].

\bibitem{Vitev:2008jh} 
  I.~Vitev,
  J.\ Phys.\ G {\bf 35}, 104011 (2008)
  [arXiv:0806.0003 [hep-ph]].

\bibitem{HorowitzPANIC2011}
  W.~A.~Horowitz,
  AIP Conf.\ Proc.\  {\bf 1441}, 889 (2012)
  [arXiv:1108.5876 [hep-ph]].

\bibitem{Buzzatti:2012pe} 
  A.~Buzzatti and M.~Gyulassy,
  arXiv:1207.6020 [hep-ph].
  
\bibitem{He:2011qa} 
  M.~He, R.~J.~Fries and R.~Rapp,
  Phys.\ Rev.\ C {\bf 86}, 014903 (2012)
  [arXiv:1106.6006 [nucl-th]].

\bibitem{Chatrchyan:2012lxa} 
  S.~Chatrchyan {\it et al.}  [CMS Collaboration],
  Phys.\ Rev.\ Lett.\  {\bf 109}, 222301 (2012)
  [arXiv:1208.2826 [nucl-ex]].
 
 \bibitem{Affolder:1999wm} 
  T.~Affolder {\it et al.}  [CDF Collaboration],
  Phys.\ Rev.\ Lett.\  {\bf 84}, 2094 (2000)
  [hep-ex/9910025].
  
\bibitem{Aaij:2012se} 
  R.~Aaij {\it et al.}  [LHCb Collaboration],
  JHEP {\bf 1211}, 031 (2012)
  [arXiv:1209.0282 [hep-ex]].
  
    
\bibitem{Strickland:2011aa} 
  M.~Strickland and D.~Bazow,
  Nucl.\ Phys.\ A {\bf 879}, 25 (2012).
  [arXiv:1112.2761 [nucl-th]].


\bibitem{Emerick:2011xu} 
  A.~Emerick, X.~Zhao and R.~Rapp,
  Eur.\ Phys.\ J.\ A {\bf 48}, 72 (2012)
  [arXiv:1111.6537 [hep-ph]].
 
\bibitem{Reed:2011zza} 
  R.~Reed [STAR Collaboration],
  J.\ Phys.\ Conf.\ Ser.\  {\bf 270}, 012026 (2011)
  [Nucl.\ Phys.\ A {\bf 855}, 440 (2011)].
  
 \end{thebibliography}
\end{document}